\documentclass [12pt, column, a4paper] {article}
\usepackage{graphicx}

\usepackage{amsmath}
\usepackage{amssymb}
\usepackage{graphicx}
\usepackage{bm}
\usepackage{slashed}

\usepackage[dvips]{epsfig}

\begin{document}

\title{The role of scalar current coupling along surfaces}

\author{F.A. Barone$^{1}$, L.H.C. Borges$^{2}$, G. Flores-Hidalgo$^{1}$, H.L. Oliveira$^{1}$, W.Y.A. da Silva$^{1}$\\ \\
$^{1}$Universidade Federal de Itajub\'a, Instituto de F\'isica e Qu\'imica,\\
Av. BPS 1303, Pinheirinho, 37500-903, Itajub\'a, MG, Brazil.\\ \\
$^{2}$Departamento de F\'\i sica, Universidade Federal de Lavras,\\
Caixa Postal 3037, 37200-900 Lavras-MG, Brazil.\\ }

%\date{}

\maketitle

%%%%%%%%%
\begin{abstract}
In this paper we propose a coupling between the complex scalar field and an external Dirac delta-like planar potential. The coupling is achieved through the Klein-Gordon current normal to the plane where the potential is concentrated. The results are obtained exactly and exhibit many peculiarities. We show that a complex scalar charge does not interact with the potential, but the potential modifies the interaction between two scalar charges if they are placed on opposite sides of the planar potential. When the coupling constant between the potential and the field goes to infinity, the classical field solutions satisfy a kind of MIT boundary conditions along the plane where the potential is concentrated.
\end{abstract}
%%%%%%%%%

\baselineskip=20pt

\section{Introduction}

The presence of material boundaries and surfaces poses a challenging problem that can be addressed through various approaches. Among them, three methods stand out: establishing boundary conditions on fields, employing field models with distinct Lagrangians defined across different spatial regions, and coupling spatially localized external potentials to fields. In special, we highlight the use of external potentials in this paper.

In the context of bosonic fields, models incorporating delta-like potentials coupled to scalar or vector fields have been explored \cite{Miltonlivro, PhysRep3532001, PRD0650272015, PRD1050202013, PRD0650202014, EPJC31132014, NPB6452002, JPA44831992, JPA63912004, hepth0201150, AnnP1662017, PRD0560172017, PRD125015}. This encompasses spherical geometries \cite{PRD0850072017, PRD0450132019, PRD0850102017, PRD0850111999}, involving not only delta-like potentials but also derivatives of delta functions \cite{PRD0250282015, PRD1250072016}. Notably, this type of model finds application in describing anisotropic planar materials \cite{PRD0850102016, ForPhys16000472017} and planar boundaries with magnetoelectric properties \cite{PhysScr1050182023, NuovCim1932013, PRD0850212012}. This last topic is addressed in a distinctive manner in reference \cite{EPJC5582021}, where a coupling term similar to the Chern-Simons one is employed.

Delta-like couplings are also explored in models involving fermionic fields, as in references \cite{JPAMT38522012023,AnnPhys4422004}, just to mention a few examples.

It is important to highlight that delta functions are indeed very useful tools to describe null-range interactions \cite{JMS42}, an issue that can find a wide range of applications, even in the non-relativistic context, such as in atomic physics \cite{livrofisicaatomica} and in fermionic dynamics with external potentials \cite{FP7}, just to mention a few. In reference \cite{livromodelossoluveis}, we can find a review of this topic where delta-like potentials are treated solely in terms of boundary conditions \cite{PRA66,JMAMG26,livroperturbacaosingular,Griffiths} — an approach that can be very convenient in many problems.

Chern-Simons-type models exhibit a wide range of peculiar features with numerous applications. They are characterized by the inclusion in the Lagrangian of a term involving the first derivative of a field and the field itself. Notable examples include the Maxwell-Chern-Simons Lagrangian \cite{AnnPhys3721982, BJP13052006, RBEF33092011}, the Cremmer-Sherk-Kalb-Ramond model \cite{NPB1171974, MPLA5591991, PRD0650262011}, the Carroll-Field-Jakiew model \cite{PRD12311990, NPB2472001, NPB2142003, PRD1250112003, PRD0250052003, EPJC5112005, EPJC1272005}, among others.

Models involving scalar fields remain relevant in both theoretical and experimental contexts. The scalar field is frequently utilized to describe condensed matter systems, such as phonons and magnons. Additionally, the scalar field is commonly employed as a toy model to mimic the properties of vector fields in specific situations. 

Therefore, it is natural to question the roles of models for the scalar field involving terms with its first derivative, akin to the Chern-Simons-type term. Furthermore, exploring whether this type of model can be employed to describe material boundaries becomes an intriguing issue of investigation.

To delve into this topic, we draw inspiration from the so called MIT boundary conditions, established for fermionic fields \cite{Chodos}. For these boundary conditions, the fermionic current takes a central role, and not the field itself. In a parallel manner, we concentrate on the Klein-Gordon current for the complex scalar field in the present work.

In this paper we propose a model where the complex scalar field couples to an external potential localized along a plane. The potential is a Dirac delta-like one concentrated along a plane and the coupling occurs through the Klein-Gordon current of the field. As far as the authors know, the model exhibits many peculiarities in comparison with the ones considered in the literature up to now. It involves terms with a first derivative on the field, what resembles Chern-Simons-like models. In spite of being a quadratic model, the differential operator present in the lagrangian is different from the one obtained in the field equations. The potential does not interact with a single stationary scalar charge, but modifies the interaction between two distinct scalar charges for setups when both charges are in opposite sides of the planar potential. When the charges are placed at the same side of the potential, the interaction between them is not affected by the potential. In the limit where the coupling constant between the potential and the field goes to infinity, the classical field solutions satisfy a kind of MIT boundary conditions along the plane where the potential is concentrated.

It would be very interesting to address the proposed problem by adopting the approach presented in reference \cite{livromodelossoluveis}, which is based on the boundary conditions of the field modes. In the present work, we will not employ this procedure. We leave this issue as an open question to be explored in a future paper.

This paper is organized as follows: in Section (\ref{secaomodelo}), we propose the model and calculate the corresponding Green function exactly. In Section (\ref{secaocargas}), we examine some effects arising from the presence of the potential and external sources. Section (\ref{limiteslambdainfinito}) is dedicated to discussing the boundary conditions of classical field solutions evaluated on the plane where the potential is concentrated, in the limit where the coupling constant between the field and the potential goes to infinity. Finally, Section (\ref{secaoconclusoes}) presents some concluding remarks.

In this paper we shall use natural units, where $\hbar=c=1$, and the (3+1)-dimensional Minkowski metric $\eta^{\mu\nu}=(1,-1,-1,-1)$. We shall also define  Minkowski vectors parallel to the $x^{3}$ plane as the one $x^{\mu}_{\|}=(x^{0},x^{1},x^{2})$.

%%%%%%%%%%%%%%%%%%%%%%%%%%%%%%%%%%%
%%%%%%%%%%%%%%%%%%%%%%%%%%%%%%%%%%%%
\section{The current coupling model}
%%%%%%%%%%%%%%%%%%%%%%%%%%%%%%%%%%%
%%%%%%%%%%%%%%%%%%%%%%%%%%%%%%%%
\label{secaomodelo}

In this section we consider the complex scalar field $\phi$ coupled to an external planar potential. The coupling between the potential and the field is made by means of the Klein-Gordon current and the plane is attained by a Dirac delta function. The lagrangian of the model is given by
\begin{eqnarray}
\label{lagrangianaprincipal}
{\cal L}=\partial_{\mu} \phi \partial^{\mu} \phi^{*} -  m^{2} \phi \phi^{*} -\lambda\delta(x^{3}-a)\ n^{\mu}j_{\mu}(x)+J\phi^{*}+J^{*}\phi
\end{eqnarray}
where $n^{\mu}=(0,0,0,1)$ is a 4-vector normal to the plane $x^{3}=a$, which is the surface where the potential is concentrated, $\delta(x^{3}-a)$ stands for the Dirac delta function, $J$ is an external source and $j_{\mu}(x)$ is the Klein-Gordon current, namely,
\begin{equation}
\label{eq92}
j^{\mu} =  \dfrac{i}{2} (\phi^{*}\partial^{\mu}\phi - \phi \partial^{\mu} \phi^{*}).
\end{equation}
and $\lambda$ is a dimensionless coupling constant between the potential and the field. In fact, $\lambda$ can be defined in the range $(-\infty,\infty)$, but we shall consider just $\lambda\geq0$ once the case $\lambda\leq0$ can be attained with a simple inversion of the vector $n^{\mu}$. 

Due to the fact that (\ref{lagrangianaprincipal}) it is quadratic in the scalar field, from now on, we shall refer to the term that involves the delta function in (\ref{lagrangianaprincipal}) as a coupling between the field and an external potential, even though one could consider it as an abuse of terminology. 

In the appendix it is shown that the hamiltonian of the model (\ref{lagrangianaprincipal}) is bounded from below, thus the theory has a vacuum ground state at the quantum level.

We start by writing the complex scalar field $\phi$ in terms of its real and imaginary parts, as well as the external source $J$, as follows
\begin{eqnarray}
\label{eq46}
\phi = \dfrac{1}{\sqrt{2}}(\phi_{1} + i\phi_{2}) 
\end{eqnarray}
\begin{eqnarray}
\label{eq49}
	J = \dfrac{1}{\sqrt{2}}(J_{1} + iJ_{2}) 
\end{eqnarray}
so, the lagrangian (\ref{lagrangianaprincipal}) reads
\begin{eqnarray}
\label{lagrangianaprincipalV2}
{\cal L}=\frac{1}{2}( \partial_{\mu} \phi_{1} \partial^{\mu} \phi_{1} + \partial_{\mu} \phi_{2} \partial^{\mu} \phi_{2}) - \frac{1}{2} m^{2} (\phi_{1}^{2} 
+\phi_{2}^{2}) + (J_{1}\phi_{1} + J_{2}\phi_{2})\cr\cr
-\dfrac{\lambda}{2}\delta(x^{3}-a)n_{\mu}(\phi_{2}\partial^{\mu}\phi_{1} - \phi_{1} \partial^{\mu} \phi_{2})\ .
\end{eqnarray}

Notice that definitions (\ref{eq46}) and (\ref{eq49}) do not decouple the fields $\phi_{1}$ and $\phi_{2}$ in (\ref{lagrangianaprincipalV2}) as is done in the Klein-Gordon free theory. 
It will be convenient to rewrite the lagrangian in a matrix form. So we define the real matrix field $\Phi(x)$
\begin{equation}
\label{eq52}
	\Phi(x) = \begin{pmatrix}
		\phi_{1}(x)\\
		\phi_{2}(x)\\
	\end{pmatrix}  
\end{equation}
and the external matrix source $\mathbb{J}$, 
\begin{equation}
\label{eq53}
\mathbb{J}(x) = \begin{pmatrix}
	J_{1}(x)\\
	J_{2}(x)\\
\end{pmatrix} \ .
\end{equation}

Denoting the $2\times 2$ identity matrix by 
\begin{equation}
\label{eq54}
	\mathbb{I} = \begin{pmatrix}
		1&0\\
		0&1\\
	\end{pmatrix}\ ,
\end{equation}
defining the $2\times 2$ matrix $\stackrel{\circ}{\mathbb{I}}$ as
\begin{equation}
\label{eq95}
	\stackrel{\circ}{\mathbb{I}} =  
	\begin{pmatrix}
		0&-1\\
		1&0\\
	\end{pmatrix}\ ,
\end{equation}
with the aid of (\ref{eq52}), (\ref{eq53}), (\ref{eq54}) and (\ref{eq95}), we can write the lagrangian (\ref{lagrangianaprincipalV2}) in the form
\begin{eqnarray}
\label{eq98}
	{\cal L}= -\frac{1}{2} \Phi^{t} \Big[(\partial_{\nu} \partial^{\nu}+m^{2}) \mathbb{I} +\lambda\delta(x^{3}-a)\stackrel{\circ}{\mathbb{I}}
\ n_{\mu}\partial^{\mu}\Big]\Phi+\mathbb{J}^{t}\Phi\ .
\end{eqnarray}

From (\ref{eq98}) we can find the dynamical equations for $\phi_{1}$ and $\phi_{2}$ (and equivalently for $\Phi$). All the dynamical equations are equivalent to  
\begin{eqnarray}
\label{eq99}
		\Big[\mathbb{I}(\partial_{\mu} \partial^{\mu} + m^{2}) + \lambda \stackrel{\circ}{\mathbb{I}} \delta(x^{3} - a) \partial_{3}+
\frac{\lambda}{2} \stackrel{\circ}{\mathbb{I}} \partial_{3}[\delta(x^{3} - a)] \Big]\Phi= \mathbb{J}\ .
\end{eqnarray}

Here, some comments are in order. The coupling term between the field and the external potential involves both the field itself and its first derivative. Furthermore, this coupling mixes the degrees of freedom of the theory, resembling Chern-Simons-like coupling for gauge fields to some extent. Additionally, the coupling term results in the differential operator in the dynamical equation for $\Phi$, on the left-hand side of the equation (\ref{eq99}), being different from the one present in the Lagrangian (\ref{eq98}). As far as we know, this feature of the proposed model is unique for quadratic theories and leads to some interesting peculiarities, as we shall see in the next section.
 
For future convenience, we define the differential operator that appears in (\ref{eq99}) as
\begin{equation}
\label{defmathbbO}
\mathbb{O}(x)=\lambda \stackrel{\circ}{\mathbb{I}} \delta(x^{3} - a) \partial_{3}+
\frac{\lambda}{2} \stackrel{\circ}{\mathbb{I}} [\partial_{3}\delta(x^{3} - a)]\ .
\end{equation}

The Green function of the system, $\mathbb{G}(x,y)$, is associated with the field dynamical equation and must exhibit a matrix structure. In fact, it satisfy the  differential equation
\begin{eqnarray}
\label{eq100}
\Big[\mathbb{I}(\partial_{\mu} \partial^{\mu} + m^{2}) + \mathbb{O}(x) \Big] \ \mathbb{G}(x,y)  = \mathbb{I} \ \delta^{4}(x -y)\ .
\end{eqnarray}

It is immediate to verify that $\mathbb{G}(x,y)$ satisfies the integral equation
\begin{equation}
\label{eq101}
	\mathbb{G}(x,y) = \mathbb{G}_{0}(x,y) - \int d^{4}z \ \mathbb{G}(x,z)\mathbb{O}(z)  \mathbb{G}_{0}(z,y).
\end{equation}
where $\mathbb{G}_{0}(x,y)$ is the Green function associated with the free field,
\begin{equation}
\label{eq72}
	\mathbb{I}(\partial_{\mu} \partial^{\mu} + m^{2}) \mathbb{G}_{0}(x,x')  = \mathbb{I} \ \delta^{4}(x-x').
\end{equation}

The free Green function $\mathbb{G}_{0}(x,y)$ can be written as a Fourier integral, as follows,
\begin{eqnarray}
\label{eq73}
\mathbb{G}_{0}(x, y) &=& \int \dfrac{d^{4}p}{(2\pi)^{4}} \ \mathbb{I} \ \left[ - \frac{e^{-ip.(x-y)}}{p^{\mu}p_{\mu} - m^{2}+i\epsilon}  
\right]\nonumber\\
&=& - \int \dfrac{d^{3}p_{||}}{(2\pi^{2})^{3}}  \left[ \int \dfrac{dp^{3}}{(2\pi)} \ \mathbb{I} \  \frac{e^{-ip^{3}(x^{3}-y^{3})}}
{p^{\mu}p_{\mu} - m^{2}+i\epsilon} 
 \right] e^{-ip_{||}(x_{||}-y_{||})} \cr\cr
&=& \int \dfrac{d^{3}p_{||}}{(2\pi^{2})^{3}} {{\cal G}\!\!\!{\bf c}}_{0}(p_{||}; x^{3},y^{3})e^{-ip_{||}.(x_{||}-y_{||})}\ ,
\end{eqnarray}
where we defined the Minkowski parallel coordinates $x_{||}=(x^{0},x^{1},x^{2})$ and momentum $p_{||}=(p^{0},p^{1},p^{2})$, and 
\begin{eqnarray}
{{\cal G}\!\!\!{\bf c}}_{0}(p_{||}; x^{3},y^{3})& = &- \int \dfrac{dp^{3}}{(2\pi)} \ \mathbb{I} \  
\frac{e^{-ip^{3}(x^{3}-y^{3})}}{p^{\mu}p_{\mu} - m^{2}+i\epsilon}\ .
\nonumber\\
&=& \mathbb{I}  \dfrac{e^{-\sigma|x^{3} - y^{3}|}}{2 \sigma},
\label{eq74}
\end{eqnarray}
where we defined the function
\begin{equation}
\sigma=\sqrt{-p_{||}^{2} + m^{2}}.
\label{defsigma}
\end{equation}

Now we write the Green function $\mathbb{G}(x,y)$  in (\ref{eq100}), in a manner similar to $\mathbb{G}_0(x,y)$ in (\ref{eq73}),
\begin{eqnarray}
\label{defcalGcorrente}
\mathbb{G}(x,y)= \int \frac{d^{3}p_{||}}{(2\pi)^{3}} {{\cal G}\!\!\!{\bf c}}(p_{||}; x^{3}, y^{3})  e^{-ip_{||}.(x_{||} - y_{||})} 
\end{eqnarray}
where ${{\cal G}\!\!\!{\bf c}}(p_{||}; x^{3}, y^{3})$ is a matrix function that must be determined.

The operator on the left hand side of (\ref{eq100}) breaks the translational invariance along the $x^{3}$ coordinate but keeps the translational invariance along the parallel coordinates, $x_{||}$. The Green function $\mathbb{G}(x,y)$ must exhibit the same translational properties of its related operator, so one can justify the Fourier structure for $\mathbb{G}(x,y)$ in (\ref{defcalGcorrente}).

Substituting (\ref{defcalGcorrente}) and (\ref{eq73}) in (\ref{eq101}) we have
\begin{eqnarray}
\label{wsx1}
\int \frac{d^{3}p_{||}}{(2\pi)^{3}} {{\cal G}\!\!\!{\bf c}}(p_{||}; x^{3}, y^{3})  e^{-ip_{||}(x_{||} - y_{||})}=
 \int \frac{d^{3}p_{||}}{(2\pi)^{3}} {{\cal G}\!\!\!{\bf c}}_{0}(p_{||}; x^{3}, y^{3})  e^{-ip_{||}(x_{||} - y_{||})}\cr\cr
-\int  d^{4}z\ \int \frac{d^{3}p_{||}}{(2\pi)^{3}} {{\cal G}\!\!\!{\bf c}}(p_{||}; x^{3}, z^{3})e^{-ip_{||}(x_{||} -
 z_{||})}\mathbb{O}(z)\int\frac{d^{3}q_{||}}{(2\pi)^{3}} {{\cal G}\!\!\!{\bf c}}_{0}(q_{||}; z^{3}, y^{3})e^{-iq_{||}(z_{||} - y_{||})}\cr
\ .
\end{eqnarray}
Integrating in above expression in  $z$ and $q_{||}$, and using (\ref{defmathbbO}) we arrive at
\begin{eqnarray}
\label{eq102}
{{\cal G}\!\!\!{\bf c}}(p_{||}; x^{3}, y^{3})= {{\cal G}\!\!\!{\bf c}}_{0}(p_{||};x^{3},y^{3})
-\dfrac{\lambda}{2}{{\cal G}\!\!\!{\bf c}}(p_{||}; x^{3}, a)\stackrel{\circ}{\mathbb{I}}\frac{\partial {{\cal G}\!\!\!{\bf c}}_{0}(p_{||};a,y^{3})}{\partial a}
+\dfrac{\lambda}{2}\frac{\partial{{\cal G}\!\!\!{\bf c}}(p_{||};x^{3},a)}{\partial a}\stackrel{\circ}{\mathbb{I}} 
{{\cal G}\!\!\!{\bf c}}_{0}(p_{||};a,y^{3})\ .
\end{eqnarray}

Now, we take $y^{3}=a$ in equation (\ref{eq102}) and in its first derivative, using (\ref{eq74}) we have 
\begin{eqnarray}
\label{edc3}
{{\cal G}\!\!\!{\bf c}}(p_{||}; x^{3},a)= {{\cal G}\!\!\!{\bf c}}_{0}(p_{||};x^{3},a)
+\frac{\lambda}{4\sigma}\frac{\partial{{\cal G}\!\!\!{\bf c}}(p_{||};x^{3},a)}{\partial a}\stackrel{\circ}{\mathbb{I}}
\end{eqnarray}
and
\begin{eqnarray}
\label{edc3b}
\frac{\partial{{\cal G}\!\!\!{\bf c}}(p_{||}; x^{3},a)}{\partial a}=\frac{\partial  {{\cal G}\!\!\!{\bf c}}_{0}
(p_{||};x^{3},a)}{\partial a}
+\frac{\lambda}{4}\sigma{{\cal G}\!\!\!{\bf c}}(p_{||}; x^{3},a)\stackrel{\circ}{\mathbb{I}}
\end{eqnarray}
From  (\ref{edc3}) and (\ref{edc3b}) it is easy to solve for ${{\cal G}\!\!\!{\bf c}}(p_{||}; x^{3},a)$ and  
${\partial{{\cal G}\!\!\!{\bf c}}(p_{||}; x^{3},a)}/{\partial a}$. We get
\begin{equation}
\label{eq104a}
{{\cal G}\!\!\!{\bf c}}(p_{||}; x^{3},a)=\frac{1}{1+\frac{\lambda^{2}}{16}}\Bigg[  {{\cal G}\!\!\!{\bf c}}_{0}(p_{||};x^{3},a)
+\frac{\lambda}{4\sigma}\frac{\partial  {{\cal G}\!\!\!{\bf c}}_{0}(p_{||};x^{3},a)}{\partial a}\stackrel{\circ}{\mathbb{I}}\Bigg]\ .
\end{equation}
and
\begin{equation}
\label{eq104b}
\frac{\partial{{\cal G}\!\!\!{\bf c}}(p_{||}; x^{3},a)}{\partial a}=\frac{16}{16+\lambda^{2}}
\Bigg[\frac{\partial{{\cal G}\!\!\!{\bf c}}_{0}(p_{||};x^{3},a)}{\partial a}+
\frac{\lambda}{4}\sigma {{\cal G}\!\!\!{\bf c}}_{0}(p_{||};x^{3},a)\stackrel{\circ}{\mathbb{I}}\Bigg]\ .
\end{equation}

As a last step to get ${{\cal G}\!\!\!{\bf c}}(p_{||}; x^{3},a)$, we substitute (\ref{eq104a}) and (\ref{eq104b}) in (\ref{eq102}) and perform some simple manipulations, resulting in
\begin{eqnarray}
\label{edc5}
{{\cal G}\!\!\!{\bf c}}(p_{||}; x^{3}, y^{3})&=&{{\cal G}\!\!\!{\bf c}}_{0}(p_{||};x^{3},y^{3})
-\frac{8\lambda}{16+\lambda^{2}}\Bigg[\frac{\lambda}{4}\Bigg(\sigma {{\cal G}\!\!\!{\bf c}}_{0} (p_{||};x^{3},a){\cal G}_{0}(p_{||};a,y^{3})
-\frac{1}{\sigma}\frac{\partial{{\cal G}\!\!\!{\bf c}}_{0}(p_{||};x^{3},a)}{\partial a}\frac{\partial{{\cal G}\!\!\!{\bf c}}_{0}(p_{||};a,y^{3})}
{\partial a}\Bigg)\nonumber\\
&~&+\stackrel{\circ}{\mathbb{I}}\Bigg({{\cal G}\!\!\!{\bf c}}_{0}(p_{||};x^{3},a)\frac{\partial{{\cal G}\!\!\!{\bf c}}_{0}(p_{||};a,y^{3})}{\partial a}-
\frac{\partial{{\cal G}\!\!\!{\bf c}}_{0}(p_{||};x^{3},a)}{\partial a}{{\cal G}\!\!\!{\bf c}}_{0}(p_{||};a,y^{3})\Bigg)\Bigg]\ .
\end{eqnarray}

Notice that expresion (\ref{edc5}) contains only the Fourier transform in the parallel coordinates of the Green function for the free  field, ${\cal G}_{0}(p_{||};x^{3},a)$, and its derivatives. With the aid of (\ref{eq74}) we can explicitly express ${{\cal G}\!\!\!{\bf c}}(p_{||}; x^{3},a)$,
\begin{eqnarray}
\label{eq110}
{{\cal G}\!\!\!{\bf c}}(p_{||}; x^{3}, y^{3})&=&\left(\dfrac{e^{-\sigma|x^{3} - y^{3}|}}{2 \sigma}
-\frac{\lambda^{2}}{(16+\lambda^{2})}\dfrac{e^{-\sigma(|x^{3} - a|+|y^{3} - a|)}}{2 \sigma}
\Big[1-{\rm sgn}(a-x^{3}){\rm sgn}(a-y^{3})\Big]\right)\mathbb{I}\nonumber\\
&~&+\frac{2\lambda}{(16+\lambda^{2})}
\dfrac{e^{-\sigma(|x^{3} - a|+|y^{3} - a|)}}{ \sigma}
\Big[{\rm sgn}(a-y^{3})-{\rm sgn}(a-x^{3})\Big]\stackrel{\circ}{\mathbb{I}}     \ .
\end{eqnarray}
with ${\rm sgn}(x)$ standing for the signal function.
%\begin{eqnarray}
%\label{defsgn}
%{\rm sgn}(x)&=&1\ \ ,\ \ x>0\cr
%{\rm sgn}(x)&=&0\ \ ,\ \ x=0\cr
%{\rm sgn}(x)&=&-1\ \ ,\ \ x<0\ .
%\end{eqnarray}

%%%%%%%%%%%%%%%%%%%%%%%%%%%%%%%%%%%%%%%
%%%%%%%%%%%%%%%%%%%%%%%%%%%%%%%%%%%%%%%
\section{The presence of stationary sources}
\label{secaocargas}

In this section, we investigate the interaction that arises from the presence of the current potential and stationary external field sources.
The Hamiltonian density associated with the Lagrangian (\ref{lagrangianaprincipal}) is
\begin{eqnarray}
\label{hamiltoniana}
{\cal H}&=&
(\partial_{0}\phi_{1})\partial_{0}\Bigg(\frac{\partial{\cal L}}{\partial(\partial_{0}\phi_{1})}\Bigg)
+(\partial_{0}\phi_{2})\partial_{0}\Bigg(\frac{\partial{\cal L}}{\partial(\partial_{0}\phi_{2})}\Bigg)-{\cal L}\cr\cr
&=&-\Phi^{t}\partial_{0}^{2}\Phi-\mathbb{J}^{t}\Phi
+\frac{1}{2}\Phi^{t}\Big[\partial_{\mu}\partial^{\mu}+m^{2}+\frac{\lambda^{2}}{m}\delta(x^{3}-a)\stackrel{\circ}{\mathbb{I}}n_{\mu}
\partial^{\mu}\Big]\Phi
\end{eqnarray}
with the corresponding energy
\begin{eqnarray}
\label{energiageral}
E
&=&\int d^{3}{\bf x}
\frac{1}{2}\Phi^{t}\Big[\partial_{\mu}\partial^{\mu}+m^{2}+\frac{\lambda^{2}}{m}\delta(x^{3}-a)\stackrel{\circ}{\mathbb{I}}
n_{\mu}\partial^{\mu}\Big]\Phi
-\int d^{3}{\bf x} \Phi^{t}\partial_{0}^{2}\Phi-\int d^{3}{\bf x} \mathbb{J}^{t}\Phi\ .
\end{eqnarray}

The field solutions to equation (\ref{eq99}) are given by
\begin{eqnarray}
\label{PhiintegralG}
\Phi(x)&=&\int d^{4}y \mathbb{G}(x,y)\mathbb{J}(y)\cr\cr
\Phi^{t}(x)&=&\int d^{4}y \mathbb{J}^{t}(z)\mathbb{G}(z,x)\ ,
\end{eqnarray}
where the second equation above is valid once $\mathbb{G}^{t}(x,z)=\mathbb{G}(z,x)$, what can be seen from (\ref{eq110}) and using the property $\stackrel{\circ}{\mathbb{I}}^{\ t}=-\stackrel{\circ}{\mathbb{I}}$.

Inserting (\ref{PhiintegralG}) in (\ref{energiageral}), using equation (\ref{eq100}) and the fact that $n_{\mu}=(0,0,0,-1)$, and performing some simple manipulations, we have
\begin{eqnarray}
\label{rfv3intermediario}
E&=&-\int d^{3}{\bf x}\Phi^{t}\partial_{0}^{2}\Phi
-\frac{1}{2}\int d^{3}{\bf x}\int d^{4}y\mathbb{J}^{t}(y)\mathbb{G}(y,x)\mathbb{J}(x)\cr\cr
&-&\frac{\lambda}{4}\int d^{3}{\bf x}\int d^{4}y\int d^{4}z
\mathbb{J}^{t}(z)\mathbb{G}(z,x)
\stackrel{\circ}{\mathbb{I}}
\mathbb{G}(x,y)\mathbb{J}(y)\Big(\partial_{3}\delta(x^{3}-a)\Big)\ . 
\end{eqnarray}

That the last term in (\ref{rfv3intermediario}) vanishes. This fact can be shown by following four steps: 1) transposing its integrand, 2) using the fact that $\stackrel{\circ}{\mathbb{I}}^{t}=-\stackrel{\circ}{\mathbb{I}}$, 3) renaming the integration variables $y\to z$ and $z\to y$, and 4) using the property of the Green function $\mathbb{G}^{t}(y,x)=\mathbb{G}(x,y)$ (which can be demonstrated with the aid of (\ref{defsigma}), (\ref{defcalGcorrente}) and (\ref{eq110})), as follows
\begin{eqnarray}
\label{rfv3intermediario1}
\int d^{4}y\int d^{4}z
\mathbb{J}^{t}(z)\mathbb{G}(z,x)
\stackrel{\circ}{\mathbb{I}}
\mathbb{G}(x,y)\mathbb{J}(y)\cr\cr
=\int d^{4}y\int d^{4}z
\mathbb{J}^{t}(y)\mathbb{G}^{t}(x,y)
\stackrel{\circ}{\mathbb{I}}^{t}\!
\mathbb{G}^{t}(z,x)\mathbb{J}(z)\cr\cr
=\int d^{4}y\int d^{4}z
\mathbb{J}^{t}(z)\mathbb{G}(z,x)
\stackrel{\circ}{\mathbb{I}}
\mathbb{G}(x,y)\mathbb{J}(y)\cr\cr
=-\int d^{4}y\int d^{4}z
\mathbb{J}^{t}(z)\mathbb{G}^{t}(x,z)
\stackrel{\circ}{\mathbb{I}}
\mathbb{G}^{t}(y,x)\mathbb{J}(y)\cr\cr
=-\int d^{4}y\int d^{4}z
\mathbb{J}^{t}(z)\mathbb{G}(z,x)
\stackrel{\circ}{\mathbb{I}}
\mathbb{G}(x,y)\mathbb{J}(y)\ .
\end{eqnarray}

The left-hand side of (\ref{rfv3intermediario1}) is the negative of the right-hand side of the same equation. Therefore, it must be equal to zero. This proves that the last term in (\ref{rfv3intermediario}) vanishes, so
\begin{eqnarray}
\label{rfv3}
E&=&-\int d^{3}{\bf x}\Phi^{t}\partial_{0}^{2}\Phi
-\frac{1}{2}\int d^{3}{\bf x}\int d^{4}y\mathbb{J}^{t}(y)\mathbb{G}(y,x)\mathbb{J}(x)\ . 
\end{eqnarray}
     
From now on, we shall restrict to the case of stationary sources, namely, $\mathbb{J}(x)=\mathbb{J}({\bf x})$. In this case, the field solution (\ref{PhiintegralG}) is also stationary, $\Phi({\bf x})$, and the first contribution on the right hand side of (\ref{rfv3}) vanishes. So, in this case, substituting (\ref{defcalGcorrente}) into (\ref{rfv3}), integrating over the time variables, and subsequently over the temporal momentum coordinates, we arrive at
\begin{eqnarray}
\label{rfv5}
E&=&-\frac{1}{2}\int d^{3}{\bf x}\int d^{3}{\bf y}\int\frac{d^{2}{\bf p}_{||}}{(2\pi)^{2}}e^{i{\bf p}_{||}.({\bf x}_{||}-{\bf y}_{||})}
\mathbb{J}^{t}({\bf x})
{{\cal G}\!\!\!{\bf c}}(p^{0}=0,{\bf p}_{||}; x^{3}, y^{3})
\mathbb{J}({\bf y})\ 
\end{eqnarray}

Substituting (\ref{eq110}) in (\ref{rfv5}) we are taken to a contribution coming from the unit matrix in (\ref{eq110}) and a contribution coming from the $\stackrel{\circ}{\mathbb{I}}$ matrix,
\begin{eqnarray}
&&\!\!\!\!\!\!\!\!
E=\int d^{3}{\bf x}d^{3}{\bf y}\int\frac{d^{2}{\bf p}_{||}}{(2\pi)^{2}}e^{i{\bf p}_{||}.({\bf x}_{||}-{\bf y}_{||})}\mathbb{J}^{t}
({\bf x})\Bigg[ -\frac{e^{-\sqrt{m^2+{\bf p}^2_{\|}}|x^{3}-y^{3}|}}{4\sqrt{m^2+{\bf p}^2_{\|}}}\mathbb{I}\nonumber\\
&&+\frac{\lambda}{4(16+\lambda^{2})}\frac{e^{-\sqrt{m^{2}+{\bf p}^{2}_{||}}(|x^{3}-a|+|y^{3}-a|)}}{\sqrt{m^{2}
+{\bf p}_{||}^{2}}}
 \Bigg(\mathbb{I}\lambda
\Big(1-{\rm sgn}(a-x^{3}){\rm sgn}(a-y^{3})\Big)\nonumber\\
&&-4\stackrel{\circ}{\mathbb{I}}\Big({\rm sgn}(a-y^{3})-{\rm sgn}(a-x^{3})\Big)\Bigg)\Bigg]\mathbb{J}({\bf y}).
\label{energiageralcorrente}
\end{eqnarray}

%%%%%%%%%%%%%%%%%%%%%%%%%%%%%%%%
%%%%%%%%%%%%%%%%%%%%%%%%%%%%%%%%%
\subsection{One single scalar charge}
\label{secaoumacarga}

The expression (\ref{energiageralcorrente}) gives the energy of the system in the presence of the current-like potential and an external stationary source.
Let us consider an external source related to what would be a point-like stationary charge for the Maxwell field. 
This type of source can be considered as a complex scalar charge, and it is characterized by
\begin{equation}
\label{eq111}
	\mathbb{J} = \mathbb{Q} \ \delta^{3}(\mathbf{x} - \mathbf{A})
\end{equation}
where $\mathbb{Q}$ is a column matrix and $\mathbf{A}$ is the position where the scalar charge is placed. For convenience, we write the matrix
 $\mathbb{Q}$ in the form
\begin{equation}\label{eq112}
\mathbb{Q} = \begin{pmatrix}
	q_{1}\\
	q_{2}\\
\end{pmatrix} \ .
\end{equation}

The first term on the right hand side of (\ref{energiageralcorrente}) gives a divergent contribution which corresponds to the self-energy of the source. This divergence is present even in the absence of the potential and does not depend on the distance between the source and the plane where the potential is concentrated.

The interaction energy between the source and the potential is given by the second and third lines of equation (\ref{energiageralcorrente}). From now on, with no lost of generality, we shall take a coordinate system where the vector ${\bf A}$ has its only non-zero component as the normal one to the plane where the external potential is concentrated, namely, ${\bf A}=(0,0,A)$. For such a choice, we can see that the contribution coming from the first term of the second line of (\ref{energiageralcorrente}) vanishes once, due to the Dirac delta function, it is proportional to the factor $1-{\rm sgn}(a-A){\rm sgn}(a-A)=0$. Also, the contribution coming from the third line of equation (\ref{energiageralcorrente}) is equal to zero due to the fact that it is proportional to 
\begin{eqnarray}
(q_{1}\ \ q_{2})\stackrel{\circ}{\mathbb{I}}\begin{pmatrix}
	q_{1}\\
	q_{2}\\
\end{pmatrix}=0\ ,
\end{eqnarray}
what can be seen from definition (\ref{eq95}). Besides, this fourth contribution is proportional to the factor ${\rm sgn}(a-A)-{\rm sgn}(a-A)=0$.

Therefore, there is no interaction between the current-like potential and the point-like delta source. To the best of our knowledge, this is the first field model where a spatially localized potential does not interact with an external charge source.

%%%%%%%%%%%%%%%%%%%%%%%%%%%%%%%%%%
%%%%%%%%%%%%%%%%%%%%%%%%%%%%%%%
\subsection{Two scalar charges and the current potential}

In this subsection, we consider a system composed of two point-like stationary scalar charges, each described by the equation (\ref{eq111}), and the current-like potential given in the model (\ref{lagrangianaprincipal}). The total source is composed of the sum of two contributions, one from each point-like charge, as follows
\begin{eqnarray}
\label{eq122}
	\mathbb{J} = \mathbb{J}_{B}+\mathbb{J}_{C}=\mathbb{Q}_{B} \ \delta^{3}(\mathbf{x} - \mathbf{B}) +\mathbb{Q}_{C} \ \delta^{3}(\mathbf{x} 
- \mathbf{C}),
\end{eqnarray}
where we defined
\begin{eqnarray}
	\mathbb{J}_{B}(x) = \mathbb{Q}_{B} \ \delta^{3}(\mathbf{x} - \mathbf{B})\cr\cr
	\mathbb{J}_{C}(x) = \mathbb{Q}_{C} \ \delta^{3}(\mathbf{x} - \mathbf{C})
\end{eqnarray}
corresponding to the sources for each charge, placed at ${\bf B}$ and ${\bf C}$.

When we substitute the total source (\ref{eq122}) in (\ref{energiageralcorrente}) we have the direct terms and the exchange terms. The direct terms encompass only contributions solely from $\mathbb{J}_{B}$ and contributions solely from $\mathbb{J}_{C}$ (as well as from the external potential). 
From the direct terms we obtain the self energies of the scalar charges $B$ and $C$, in addition to the the interaction energies between the charge $B$ with the potential and charge $C$ with the potential.
As discussed in the single point charge case treated above, the self-energies are disregarded, as we are interested solely in the interactions between the point charges and the external potential, as well as between the charges themselves.
The terms which account for the interaction of each charge, solely, with the potential vanish, as we have seen in section (\ref{secaoumacarga}).
Therefore, the direct terms do not contribute for the interaction energy of the system.

So, let us focus on the exchange terms, namely, the ones that involve simultaneously both $\mathbb{J}_{B}$ and $\mathbb{J}_{C}$. 
The contribution coming from the first line of (\ref{energiageralcorrente}), taking only the exchange terms, is given by the integral
\begin{eqnarray}
\label{edc4}
E_{0,INT}=-\int d^{3}{\bf x}d^{3}{\bf y}\int\frac{d^{2}{\bf p}_{||}}{(2\pi)^{2}}e^{i{\bf p}_{||}.({\bf x}_{||}-{\bf y}_{||})}
\frac{e^{-\sqrt{m^2+{\bf p}^2_{\|}}|x^{3}-y^{3}|}}{4\sqrt{m^2+{\bf p}^2_{\|}}}\cr\cr
\Bigg[(q_{1B},q_{2B})\delta^{3}(\mathbf{x} - \mathbf{B})
\begin{pmatrix}
	q_{1C}\\
	q_{2C}\\
\end{pmatrix}\delta^{3}(\mathbf{y} - \mathbf{C})\cr\cr
+(q_{1C},q_{2C})\delta^{3}(\mathbf{x} - \mathbf{C})
\begin{pmatrix}
	q_{1B}\\
	q_{2B}\\
\end{pmatrix}\delta^{3}(\mathbf{y} - \mathbf{B})
\Bigg]\ .
\end{eqnarray}

Integrating (\ref{edc4}) over the spatial coordinates, defining the vectors parallel to the plane where the potential is concentrated, ${\bf A}_{||}=(A^{1},A^{2},0)$ 
and ${\bf B}_{||}=(B^{1},B^{2},0)$, using polar coordinates for the parallel momentum ${\bf p}_{||}$, with $r$ and $\phi$ standing for the radial and angular 
coordinates, respectively, and performing some manipulations, we can write
\begin{eqnarray}
\label{rfv6}
E_{0,INT}=-(q_{1B}q_{1C}+q_{2B}q_{2C})\frac{1}{2}\frac{1}{(2\pi)^{2}}\int_{0}^{\infty}dr\ r \frac{e^{-\sqrt{m^2+r^2}|B^{3}-C^{3}|}}{\sqrt{m^2+r^2}}
\int_{0}^{2\pi}d\phi e^{ir|{\bf B}_{||}-{\bf C}_{||}|\cos(\phi)}\ .
\end{eqnarray}

The integral above is a representation of the Bessel function $J_{0}$, 
\begin{equation}
\label{BesselJ0}
J_{0}(x)=\frac{1}{2\pi}\int_{0}^{2\pi}d\phi\ e^{ix\cos(\phi)}\ ,
\end{equation}
so we can write
\begin{eqnarray}
\label{edc5a}
E_{0,INT}
=-(q_{1B}q_{1C}+q_{2B}q_{2C})\frac{m}{4\pi}\int_{1}^{\infty}du\ e^{-um|B^{3}-C^{3}|}J_{0}\big(m\sqrt{u^{2}-1}|{\bf B}_{||}-{\bf C}_{||}|\big) \ ,
\end{eqnarray}
where we performed the change in the integration variable $u=\sqrt{1+\frac{r^{2}}{m^{2}}}$.

Using the formula M0179a of reference \cite{Gradshestein}, namely,
\begin{equation}
\int_{1}^{\infty}dx\ e^{-\alpha x}J_{0}(\beta\sqrt{x^{2}-1})=\frac{1}{\sqrt{\alpha^{2}+\beta^{2}}}e^{-\sqrt{\alpha^{2}+\beta^{2}}}
\end{equation}
for the integral in (\ref{edc5}), we get
\begin{eqnarray}
E_{0,INT}&=&-(q_{1B}q_{1C}+q_{2B}q_{2C})\frac{1}{4\pi}\sqrt{\frac{2}{\pi}}\big( m|{\bf B}-{\bf C}|\Big)^{-1/2}
K_{1/2}\big(m|{\bf B}-{\bf C}|\big)\nonumber\\
\label{E0final}
&=&-(q_{1B}q_{1C}+q_{2B}q_{2C})\frac{1}{4\pi}\frac{e^{-m|{\bf B}-{\bf C}|}}{|{\bf B}-{\bf C}|}\ .
\end{eqnarray}

Notice that (\ref{E0final}) is nothing else than the Yukawa potential between the charges $B$ and $C$. This result is expected because this contribution is obtained just from the free Green function (in the absence of the potential).

The next contribution for the interaction energy comes from the second line of (\ref{energiageralcorrente}),
\begin{eqnarray}
E_{1,INT}=\frac{\lambda^{2}}{4(16+\lambda^{2})}
\int d^{3}{\bf x}d^{3}{\bf y}\int\frac{d^{2}{\bf p}_{||}}{(2\pi)^{2}}e^{i{\bf p}_{||}.({\bf x}_{||}-{\bf y}_{||})}\cr\cr
\frac{e^{-\sqrt{m^{2}+{\bf p}^{2}_{||}}(|x^{3}-a|+|y^{3}-a|)}}{\sqrt{m^{2}+{\bf p}_{||}^{2}}}
\Big(1-{\rm sgn}(a-x^{3}){\rm sgn}(a-y^{3})\Big)\cr\cr
\times \Bigg[(q_{1B}, q_{2B})\delta^{3}(\mathbf{x} - \mathbf{B})
\begin{pmatrix}
	q_{1C}\\
	q_{2C}\\
\end{pmatrix}\delta^{3}(\mathbf{y} - \mathbf{C})\cr\cr
+(q_{1C},q_{2C})\delta^{3}(\mathbf{x} - \mathbf{C})
\begin{pmatrix}
	q_{1B}\\
	q_{2B}\\
\end{pmatrix}\delta^{3}(\mathbf{y} - \mathbf{B})\ .
\Bigg]
 \end{eqnarray}

Now we proceed in a similar way we have made to obtain (\ref{E0final}). The result is
\begin{eqnarray}
\label{E1final}
E_{1,INT}=\frac{\lambda^{2}}{4\pi(16+\lambda^{2})}
\big(q_{1B}q_{1C}+q_{2B}q_{2C}\big)\Big(1-{\rm sgn}(a-B^{3}){\rm sgn}(a-C^{3})\Big)\cr\cr
\times
\frac{e^{-m\sqrt{(|B^{3}-a|+|C^{3}-a|)^{2}+|{\bf B}_{||}-{\bf C}_{||}|^{2}}}}{\sqrt{(|B^{3}-a|+
|C^{3}-a|)^{2}+|{\bf B}_{||}-{\bf C}_{||}|^{2}}}\ .
\end{eqnarray}

The contribution (\ref{E1final}) takes into account the presence of both the potential and the two charges simultaneously. It vanishes when both charges are  at the same side of the plane where the potential is concentrated, due to the fact that the factor $\Big(1-{\rm sgn}(a-B^{3}){\rm sgn}(a-C^{3})\Big)$ equals to zero in  this case. The contribution also vanishes when $\lambda=0$, what is expected because in this case there is no coupling between the field and the 
potential.

The last contribution to the interaction energy is obtained from the third line of equation (\ref{energiageralcorrente}). Pursuing as with the other terms, we obtain
\begin{eqnarray}
E_{2,INT}&=&\frac{\lambda}{(16+\lambda^2)}
\int d^{3}{\bf x}d^{3}{\bf y}\int\frac{d^{2}{\bf p}_{||}}{(2\pi)^{2}}e^{i{\bf p}_{||}.({\bf x}_{||}-{\bf y}_{||})}
\frac{e^{-\sqrt{m^{2}+{\bf p}^{2}_{||}}(|x^{3}-a|+|y^{3}-a|)}}{\sqrt{m^{2}+{\bf p}_{||}^{2}}}
\Big({\rm sgn}(a-y^{3})-{\rm sgn}(a-x^{3})\Big)\nonumber\\
& &\times \Bigg[(q_{1B},q_{2B})\stackrel{\circ}{\mathbb{I}}
\begin{pmatrix}
	q_{1C}\\
	q_{2C}\\
\end{pmatrix}\delta^{3}(\mathbf{x} - \mathbf{B})\delta^{3}(\mathbf{y} - \mathbf{C})
+(q_{1C},q_{2C})\stackrel{\circ}{\mathbb{I}}
\begin{pmatrix}
	q_{1B}\\
	q_{2B}\\
\end{pmatrix}\delta^{3}(\mathbf{x} - \mathbf{C})\delta^{3}(\mathbf{y} - \mathbf{B})\Bigg]\nonumber\\
&=&
\frac{1}{4\pi}\frac{4\lambda}{(16+\lambda^2)}(q_{1B}q_{2C}-q_{2B}q_{1C})\Big({\rm sgn}(a-B^{3})-{\rm sgn}(a-C^{3})\Big)\cr\cr
& &\times
\frac{e^{-m\sqrt{(|B^{3}-a|+|C^{3}-a|)^{2}+|{\bf B}_{||}-{\bf C}_{||}|^{2}}}}{\sqrt{(|B^{3}-a|+
|C^{3}-a|)^{2}+|{\bf B}_{||}-{\bf C}_{||}|^{2}}}\ .
\label{edc6}
\end{eqnarray}

Once again, the contribution (\ref{edc6}) vanishes when both scalar charges are placed on the same side of the external potential. Additionally, when the charges are positioned on the plane where the potential is concentrated, the contribution (\ref{edc6}) also vanishes.
Besides, expression (\ref{edc6}) is equal to zero when both charges are equal, namely, $q_{1B}=q_{1C}$ and $q_{2B}=q_{2C}$.

Therefore, the total interaction energy between the point charges is given by the sum of (\ref{E0final}), (\ref{E1final}) and (\ref{edc6}), 
\begin{eqnarray}
\label{Efinal}
E_{INT}
&=&\frac{1}{4\pi}(q_{1B}q_{1C}+q_{2B}q_{2C})\Bigg[-\frac{e^{-m|{\bf B}-{\bf C}|}}{|{\bf B}-{\bf C}|}\nonumber\\
&+&\frac{\lambda^{2}}{16+\lambda^{2}}\Big(1-{\rm sgn}(a-B^{3}){\rm sgn}(a-C^{3})\Big)
\frac{e^{-m\sqrt{(|B^{3}-a|+|C^{3}-a|)^{2}+|{\bf B}_{||}-{\bf C}_{||}|^{2}}}}{\sqrt{(|B^{3}-a|+|C^{3}-a|)^{2}
+|{\bf B}_{||}-{\bf C}_{||}|^{2}}}\Bigg]\cr\cr
&+&\frac{1}{4\pi}\frac{4\lambda}{(16+\lambda^2)}(q_{1B}q_{2C}-q_{2B}q_{1C})\Big({\rm sgn}(a-B^{3})-{\rm sgn}(a-C^{3})\Big)\cr\cr
& &\times
\frac{e^{-m\sqrt{(|B^{3}-a|+|C^{3}-a|)^{2}+|{\bf B}_{||}-{\bf C}_{||}|^{2}}}}{\sqrt{(|B^{3}-a|+
|C^{3}-a|)^{2}+|{\bf B}_{||}-{\bf C}_{||}|^{2}}}\ .
\end{eqnarray}

As pointed out previously, the first contribution for the energy (\ref{Efinal}) is the direct interaction between two scalar charges and it is present even in the absence of the external potential, when $\lambda=0$. The second and third contributions are induced by the presence of the potential. They vanish when the scalar charges are placed on the same side of the planar potential. When $\lambda=0$, the second and third factors in (\ref{Efinal}) vanish, as they should.

%%%%%%%%%%%%%%%%%%%%%%%%%%%%%%%%%%%%%%%%%%%%%%
%%%%%%%%%%%%%%%%%%%%%%%%%%%%%%%%%%%%%%%%%%%%%
\subsubsection{Interpreting the result}

We can interpret the result (\ref{Efinal}) in an interesting manner. Let us take, without loss of generality, the potential set at the plane $x^{3}=0$,  which is equivalent to setting $a=0$. Besides, let us restrict to situations where the scalar charges are placed on opposite sides of the plane $x^3=0$. Otherwise, as mentioned earlier, the potential does not have any influence on the interaction between scalar charges.
In this case, we choose $B^{3}>0$ and $C^{3}<0$, which makes $1-{\rm sgn}(a-B^{3}){\rm sgn}(a-C^{3})=2$ and $\Big({\rm sgn}(a-B^{3})-{\rm sgn}(a-C^{3})\Big)=2$. Besides, we have $-C^{3}=|C^{3}|$ and $B^{3}=|B^{3}|$, so we can write
\begin{eqnarray}
\sqrt{(|B^{3}-a|+|C^{3}-a|)^{2}+|{\bf B}_{||}-{\bf C}_{||}|^{2}}=\cr
=\sqrt{(|B^{3}|+|C^{3}|)^{2}+|{\bf B}_{||}-{\bf C}_{||}|^{2}}\cr
=\sqrt{(B^{3}-C^{3})^{2}+|{\bf B}_{||}-{\bf C}_{||}|^{2}}=|{\bf B}-{\bf C}|\ ,
\end{eqnarray}
and expression (\ref{Efinal}) reads
\begin{eqnarray}
\label{Efinal2}
E_{INT}=-\frac{1}{4\pi}
\mathbb{Q}_{B}^{t}\Bigg[
\frac{16-\lambda^{2}}{16+\lambda^{2}}\mathbb{I}+\frac{8\lambda}{16+\lambda^2}\stackrel{\circ}{\mathbb{I}}
\Bigg]\mathbb{Q}_{C}
\frac{e^{-m|{\bf B}-{\bf C}|}}{|{\bf B}-{\bf C}|}\ ,
\end{eqnarray}
that exhibits a standard Yukawa behavior.

The interaction between the charges $B$ and $C$ without the presence of the planar potential is recovered by setting the coupling constant $\lambda=0$ in (\ref{Efinal2}),
\begin{equation}
\label{Esempotencial}
E_{INT}(\lambda=0)=-\frac{1}{4\pi}
\mathbb{Q}_{B}^{t}\mathbb{Q}_{C}
\frac{e^{-m|{\bf B}-{\bf C}|}}{|{\bf B}-{\bf C}|}\ .
\end{equation}

The comparison of expressions (\ref{Efinal}) and (\ref{Efinal2}) reveals that by fixing the values for the charges intensities $q_{1B}$, $q_{2B}$, $q_{1C}$ and $q_{2C}$, the presence of the potential between the charges can result in a sign inversion of the interaction between them, depending on the value of the parameter $\lambda$. Moreover, the role of the potential for the interparticle interaction is solely to introduce a modulation factor $\alpha(\lambda)$ and a mixing factor between the components of the charges intensities, $\beta(\lambda)$, defined respectively by
\begin{eqnarray}
\label{fatoresalphabeta}
\alpha(\lambda)=\frac{16-\lambda^{2}}{16+\lambda^{2}}\cr\cr
\beta(\lambda)=\frac{8\lambda}{16+\lambda^2}\ .
\end{eqnarray}

\begin{figure}[h]
\label{graficosalphabeta}
\centering
\includegraphics[scale=1.1]{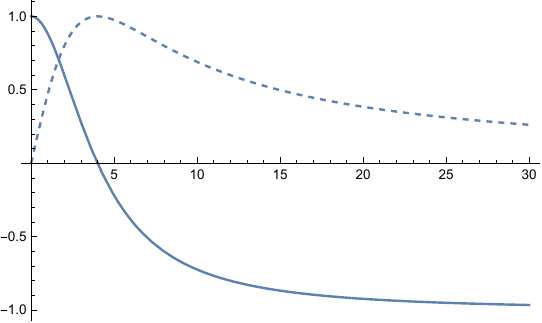}
\caption{In the solid line we have $\alpha(\lambda)$ and in the dashed line we have $\beta(\lambda)$. The horizontal axis is $\lambda$.}
\end{figure}

In Figure (\ref{graficosalphabeta}), we observe a plot for the factor $\alpha(\lambda)$ represented by a solid line and for the factor $\beta(\lambda)$ represented by a dashed line. When $\lambda=0$, we have a maximum for $\alpha(0)=1$ and a minimum for $\beta(0)=0$, and equation (\ref{Efinal2}) reduces to (\ref{Esempotencial}) as expected. The function $\alpha(\lambda)$ is positive in the interval $0<\lambda<4$, becomes zero at $\lambda=4$, and turns negative for $\lambda>4$. The function $\beta(\lambda)$ is always positive and reaches its maximum at $\lambda=4$, where $\beta(4)=1$. As $\lambda\to\infty$, we observe a minimum for $\alpha(\infty)\to-1$ and $\beta(\infty)\to0$.

For small values of $\lambda$, the contribution to the interaction energy involving the attenuation factor $\alpha(\lambda)$ dominates in comparison to the one coming from the mixing factor $\beta(\lambda)$, as well as for large values of $\lambda$, but with a sign inversion in this second case. Around the value $\lambda=4$, the contribution involving the mixing factor $\beta(\lambda)$ prevails over the one coming from the attenuation factor $\alpha(\lambda)$.

The energy (\ref{Efinal2}) is not symmetric with respect to an interchange between the particles $B$ and $C$. This interesting property is a feature of the model (\ref{lagrangianaprincipal}), which distinguishes between the two different sides of the planar potential due to the definition of the normal vector $n^{\mu}$.

From the perspective of particle $B$, located on the right side of the planar potential, the energy (\ref{Efinal2}) is the same as the one obtained with a fictitious particle placed at the same position as particle $C$, but with an effective charge given by
\begin{equation}
\mathbb{Q}_{C(eff)}=
\Bigg[
\frac{16-\lambda^{2}}{16+\lambda^{2}}\mathbb{I}+\frac{8\lambda}{16+\lambda^2}\stackrel{\circ}{\mathbb{I}}
\Bigg]\mathbb{Q}_{C}\ .
\end{equation}

Similarly, from the perspective of particle $C$, situated on the left side of the planar potential, the energy (\ref{Efinal2}) is equivalent to the one obtained with a fictitious particle at the same position as particle $B$, but with an effective charge given by
\begin{equation}
\mathbb{Q}_{B(eff)}=
\Bigg[
\frac{16-\lambda^{2}}{16+\lambda^{2}}\mathbb{I}-\frac{8\lambda}{16+\lambda^2}\stackrel{\circ}{\mathbb{I}}
\Bigg]\mathbb{Q}_{B}\ .
\end{equation}

%%%%%%%%%%%%%%%%%%%%%%%%%%%%%%%%%%%%%%%%%%%%%%
%%%%%%%%%%%%%%%%%%%%%%%%%%%%%%%%%%%%%%%%%%%%%%
\subsubsection{Charges at the potential plane}

\begin{figure}[h]
\label{noplanodopotencial}
\centering
\includegraphics[scale=0.7]{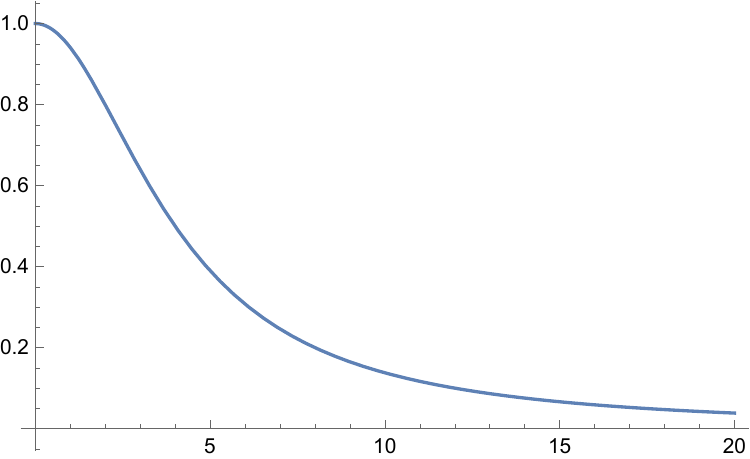}
\caption{In the vertical axis we have the factor $\frac{16}{16+\lambda^{2}}$ that modulates the dependence on $\lambda$ of the energy (\ref{Ecargasplanopotencial}). The horizontal axis is $\lambda$.}
\end{figure}

A very interesting peculiarity of the model (\ref{lagrangianaprincipal}) is the fact that the interaction energy (\ref{Efinal}) is finite when the charges  lie on the plane where the potential is concentrated. In this case, with no loss of generality, we can take a coordinates system where the charges positions are given by ${\bf B}=(R,0,a)$ (with $R>0$) and ${\bf C}=(0,0,a)$.
In this case, the energy (\ref{Efinal}) becomes
\begin{eqnarray}
\label{Ecargasplanopotencial}
E_{INT}&=&\frac{1}{4\pi}
(q_{1B}q_{1C}+q_{2B}q_{2C})\Bigg[-\frac{e^{-mR}}{R}
+\frac{\lambda^{2}}{(16+\lambda^{2})}\frac{e^{-mR}}{R}\Bigg]\nonumber\\
&=&
-\frac{4}{\pi(16+\lambda^2)}(q_{1B}q_{1C}+q_{2B}q_{2C})\frac{e^{-mR}}{R}
\ ,
\end{eqnarray}
which is exactly the same expression that we would have obtained in the absence of the potential (the Yukawa potential) multiplied by the non-negative numerical factor $0\leq\frac{16}{16+\lambda^{2}}\leq 1$. As a consequence, the influence of the potential is merely to attenuate the intensity of the interaction by that numerical factor.  We display in figure (\ref{noplanodopotencial})  that factor as function of the parameter $\lambda$. We see that in the limit $\lambda\to\infty$, the interaction (\ref{Ecargasplanopotencial}) vanishes.

The finiteness of (\ref{Ecargasplanopotencial}) is a peculiar feature of the model (\ref{lagrangianaprincipal}) because, as far as the authors know, in models where fields couple to external delta-like potentials, the interactions between external sources usually diverge when a source lies on the regions where the potentials are concentrated.

%%%%%%%%%%%%%%%%%%%%%%%%%%%
%%%%%%%%%%%%%%%%%%%%%%%%%%%
\section{The limit $\lambda\to\infty$}
\label{limiteslambdainfinito}

In this section, we direct our focus toward the field solutions of the model (\ref{lagrangianaprincipal}) when evaluated on the plane where the potential is concentrated and, particularly, in the limit as $\lambda$ approaches infinity.

Taking into account equation (\ref{PhiintegralG}), we can see that
\begin{equation}
\label{tgb1}
\Phi(x)_{x^{3}=a}=\int d^{4}y \Big[\mathbb{G}(x,y)_{x^{3}=a}\Big]\mathbb{J}(y)\ .
\end{equation}
So we must consider the Green function (\ref{defcalGcorrente}), with (\ref{eq110}) evaluated at $x^{3}=a$, 
\begin{eqnarray}
\label{tgb2}
\mathbb{G}(x,y)_{x^{3}=a}&=& \int \frac{d^{3}p_{||}}{(2\pi)^{3}} {{\cal G}\!\!\!{\bf c}}(p_{||}; a, y^{3})  e^{-ip_{||}.(x_{||} - y_{||})}\cr\cr
&=& \int \frac{d^{3}p_{||}}{(2\pi)^{3}}e^{-ip_{||}.(x_{||} - y_{||})}\dfrac{e^{-\sigma|y^{3}-a|}}{2 \sigma}
\Bigg[\frac{16\mathbb{I}+4\lambda{\rm sgn}(a-y^{3})\stackrel{\circ}{\mathbb{I}}}{16+\lambda^{2}}\Bigg]
\end{eqnarray}

In the limit $\lambda\to\infty$, the integrand in (\ref{tgb2}) equals to zero, and so does the field solution (\ref{tgb1}) at the plane $x^{3}=a$,
\begin{eqnarray}
\label{condDirichlet}
\lim_{\lambda\to\infty}\mathbb{G}(x,y)_{x^{3}=a}&=&0\cr\cr
\lim_{\lambda\to\infty}\Phi(x)_{x^{3}=a}&=&0\ .
\end{eqnarray}
It is precisely the so-called Dirichlet boundary condition for the matrix field solution $\Phi$ and its associated Green function $\mathbb{G}(x,y)$.

Let us also consider the normal derivative to the plane $x^{3}=a$ of the field solution (\ref{PhiintegralG}) evaluated at the plane $x^{3}=a$. Once again, from equations (\ref{defcalGcorrente}) and  (\ref{eq110}), we can write
\begin{equation}
\label{tgb3}
\partial_{3}\Phi(x)|_{x^{3}=a}=\int d^{4}y \frac{\partial\mathbb{G}(x,y)}{\partial x^{3}}\Big|_{x^{3}=a}\mathbb{J}(y)\ .
\end{equation}

Following similar steps employed in section (\ref{secaomodelo}), with the aid of (\ref{defcalGcorrente}), (\ref{edc5}) and (\ref{eq74}) we can show that
\begin{eqnarray}
\frac{\partial\mathbb{G}(x,y)}{\partial x^{3}}\Big|_{x^{3}=a}&=&\int \frac{d^{3}p_{||}}{(2\pi)^{3}} \frac{\partial{{\cal G}\!\!\!{\bf c}}(p_{||}; x^{3}, y^{3})}{\partial x^{3}}\Big|_{x^{3}=a} e^{-ip_{||}.(x_{||} - y_{||})}\cr\cr
&=&\int \frac{d^{3}p_{||}}{(2\pi)^{3}}e^{-ip_{||}.(x_{||} - y_{||})}\dfrac{e^{-\sigma|y^{3}-a|}}{2}\frac{\mbox{sgn}(y^{3}-a)\mathbb{I}-4\lambda\stackrel{\circ}{\mathbb{I}}}{16+\lambda^{2}}\ .
\end{eqnarray}

In the limit $\lambda\to\infty$ we have
\begin{eqnarray}
\label{condNeumann}
\lim_{\lambda\to\infty}\frac{\partial\mathbb{G}(x,y)}{\partial x^{3}}\Big|_{x^{3}=a}&=&0\cr\cr
\lim_{\lambda\to\infty}\partial_{3}\Phi(x)|_{x^{3}=a}&=&0\ ,
\end{eqnarray}
which is the so-called Neumann boundary condition, along the plane $x^{3}=a$, for the matrix field solution $\Phi$ and its associated Green function $\mathbb{G}(x,y)$.

Lastly, we consider the normal Klein-Gordon current along the plane where the potential is concentrated, given by
\begin{equation}
\label{tgb4}
n_{\mu}j^{\mu}(x)|_{x^{3}=a}=j^{3}(x)|_{x^{3}=a}=\frac{1}{2}\left[\Phi^{t}(x)\stackrel{\circ}{\mathbb{I}}\partial^{3}\Phi(x)\right]|_{x^{3}=a}\ .
\end{equation}

Due to the results (\ref{condDirichlet}) and (\ref{condNeumann}), we can see that the right hand side of (\ref{tgb4}) equals to zero, namely,
\begin{equation}
\label{condMIT}
n_{\mu}j^{\mu}(x)|_{x^{3}=a}=j^{3}(x)|_{x^{3}=a}=0\ .
\end{equation}

The result (\ref{condMIT}) is a kind of MIT boundary condition along the plane $x^{3}=a$ for the scalar field.

%%%%%%%%%%%%%%%%%%%%%%%%%%%%%%%%%%%%%%%
%%%%%%%%%%%%%%%%%%%%%%%%%%%%%%%%%%%%%%%%%%%%%%
\section{Conclusions and Final Remarks}
\label{secaoconclusoes}

In this paper we proposed a model where the complex scalar field couples to an external planar potential. The coupling is achieved through the Klein-Gordon 
current coupled to a Dirac delta function concentrated along a plane. In the appendix we have shown that the model is bounded from below.

We have  treated the equations in a matrix structure form in order to deal with  the field and the Green function  equations. In this way
we have founded, exactly, the Green function of the model. We have shown that the differential operator present in the dynamical field equations is different  from the one in the lagrangian.

We have investigated the types of interactions that may arise between the potential and stationary field sources in the model. Our findings reveal that a single complex scalar charge does not interact with the potential. However, the interaction between two complex scalar charges is modified by the presence of the potential, but only if they are positioned on opposite sides of the potential. This modification can be categorized into two contributions: a direct contribution and a mixing contribution, each modulated by a function of the coupling constant.

The direct contribution does not mix the two different components of the charges. It varies within the interval $[-1,1)$ and becomes zero for $\lambda=4$. On the other hand, the mixing contribution combines the two different components of the charges, displays a local maximum at $\lambda=4$ and varies within the interval $[1,0)$.

The interaction energy between the two charges in the presence of the potential undergoes a sign inversion for large values of the coupling constant $\lambda$ compared to situations where $\lambda$ is small.

In the scenario where the coupling constant between the potential and the field tends to infinity, we have demonstrated that the classical field solutions and the Green function of the model satisfy the Dirichlet, Neumann, and MIT boundary conditions along the plane where the potential is concentrated.

We leave open the question of treating the proposed model in terms of the boundary conditions of the associated wave functions \cite{andamento}. This approach may provide a direct means of analyzing the underlying divergent quantities of the problem and the field modes associated with the model.

We hope that the discussions presented in this paper prove to be relevant for investigations in the field of theories with internal symmetries, particularly those involving the coupling of fields with external potentials.

\ 

\textbf{Acknowledgments:}  F.A. Barone thanks to CNPq, under the grant 313426/2021-0.  H.L. Oliveira thanks to CNPq for invaluable financial support.

\appendix
\section{Appendix: the boundedness of the hamiltonian}

In this appendix, we intend to study the conditions under which the model (\ref{lagrangianaprincipal}) is bounded from below.

From the lagrangian density given by equation (\ref{lagrangianaprincipal}) we get the hamiltonian
\begin{equation}
H=\int d^3 {\bf x} \left[ |\dot{\phi}|^2+|\nabla \phi|^2+m^2|\phi|^2+\frac{i\lambda}{2}\delta(x^3-a){\bf n}.
(\phi^\star\nabla \phi-\phi\nabla\phi^\ast)\right]
\label{hamilA1}
\end{equation} 
where we discarded the linear term in the field, without loss of generality for the present purpose.

All the terms in the above expression are non-negative, except for the last one, which can be positive or negative. Consequently, the Hamiltonian above could be without a lower limit or unbounded from below.

A Hamiltonian unbounded from below at the quantum level lacks a ground state, which is an undesired and unphysical property. Let us investigate the conditions under which the Hamiltonian (\ref{hamilA1}) is bounded from below. We anticipate that for sufficiently small $\lambda$ parameters, the non-negative terms dominate over the last term in (\ref{hamilA1}), and in such regimes, the Hamiltonian will be bounded from below.

In the following, we establish this expectation quantitatively.

\subsection{A simplified case}

The term that would render the Hamiltonian (\ref{hamilA1}) negative is the one containing the Dirac delta function. Thus, we can analyze the influence of this term in a simpler yet more robust scenario. Initially, we examine the simplified situation where the delta function potential is substituted by a factor of 1. In this case, the Hamiltonian is denoted as
\begin{equation}
H_{s}=\int d^3 {\bf x} \left[ |\dot{\phi}|^2+|\nabla \phi|^2+m^2|\phi|^2+\frac{i\lambda}{2}{\bf n}.
(\phi^\star\nabla \phi-\phi\nabla\phi^\ast)\right]\ .
\label{hamilA2}
\end{equation} 
Performing a Fourier transform in the spatial coordinates,
\begin{equation}
\phi({\bf x},t)=\frac{1}{(2\pi)^{3/2}}\int d^3{\bf k} ~\tilde{\phi}_k(t) e^{i\bf{k}.{\bf x}}
\label{plane1}
\end{equation}
and replacing in (\ref{hamilA2}), we get
\begin{equation}
H_s=\int d^3{\bf k}\left( |\dot{\tilde{\phi}}_k|^2+\omega_k^2|\tilde{\phi}_k|^2
\right)
\label{hamilA3}
\end{equation}
where the frequency modes, $\omega_k$ are given by
\begin{equation}
\omega_k^2=\left({\bf k}-\frac{\lambda}{2m}{\bf n}\right)^2+m^2-\frac{\lambda^2}{4}.
\label{freqmod1}
\end{equation}
From (\ref{hamilA3}), we see that the Hamiltonian (\ref{hamilA2}) will be positively defined or bounded from below if the frequency modes satisfy $\omega_k^2 \geq 0$. From this, we obtain the condition
\begin{equation}
\frac{\lambda^2}{4}\leq m^2 .
\label{lambda1}
\end{equation}
%%%%

\subsection{The case with Dirac delta function}

In the simplified case, we could express the hamiltonian (\ref{hamilA1}) in terms of the spatial Fourier transform $\tilde{\varphi}$. However, due to the delta function along the $x^3$ coordinate, the hamiltonian will not diagonalize as in (\ref{hamilA3}), making 
the analysis very challenging.

In this case, we have to expand the field $\phi$ in terms of a complete set of functions that diagonalize the Hamiltonian (\ref{hamilA1}). Disregarding surface terms, we have, for (\ref{hamilA1}),
\begin{equation}
H=\int d^3{\bf x} \left(|\dot{\phi}|^2+\frac{1}{2}\phi^\ast D\phi+\frac{1}{2}\phi D^\ast\phi^\ast
\right)
\label{hamilA4}
\end{equation}
where $D$ is the differential operator
\begin{equation}
D=-\nabla^2+m^2+i\lambda \delta(x^3-a){\bf n}.\nabla+\frac{i\lambda}{2}{\nabla}.(\delta(x^3-a){\bf n})
\label{diffop1}
\end{equation}
and $D^\ast$ is the complex conjugate of $D$. Note that both $D$ and $D^\ast$ are Hermitian differential operators. The last term in (\ref{diffop1}) is a divergence and cancels in (\ref{hamilA4}) with its counterpart in $D^\ast$. It is crucial to define these operators in this manner; otherwise, they will not be Hermitian operators in the functional space.

Expanding the field as
\begin{equation}
\phi({\bf x},t)=\sum_s q_s(t)u_s({\bf x})
\label{modes}
\end{equation}
where $\{u_s({\bf x})\}$ are orthonormal eigenfunctions of $D$,
\begin{equation}
\left[-\nabla^2+m^2+i\lambda\delta(x^3-a){\bf n}.\nabla+\frac{i\lambda}{2}{\nabla}.(\delta(x^3-a)
{\bf n})\right]u_s({\bf x})=\omega_s^2u_s({\bf x}).
\label{eigen1}
\end{equation}
with eigenvalues $\omega_s^2$, replacing (\ref{modes}) in (\ref{hamilA4}), using (\ref{eigen1}) and the orthonormality of $\{u_s({\bf x})\}$,
we get
\begin{equation}
H=\sum_{s}\left( |\dot{q}_s|^2+\omega_s^2 |q_s|^2\right).
\label{HamilD}
\end{equation}
In that way, to set the conditions under which the above hamiltonian is bounden from below, it will be sufficient to set the conditions for the eigenvalues $\omega_s^2$ to be non negative. To simplify notation, we will henceforth use $z=x^3$.

Writing
\begin{equation}
u_s({\bf x})= e^{i{\bf k}_{||}.{\bf x}_{||}} v_s(z)\ ,
\end{equation}
replacing in (\ref{eigen1}) and using the fact that ${\bf n}=(0,0,1)$, we have
\begin{equation}
\left[ -\frac{d^2}{dz^2}+i\lambda\delta(z-a)\frac{d}{dz}
+\frac{i\lambda}{2}\delta'(z-a)
\right]v_s(z)
=
\alpha _s v_s(z).
\label{eigen4}
\end{equation}
where $\delta'(z)=d\delta(z)/dz$ and
\begin{equation}
\omega_s^2(k)={\bf k}_{||}^2+m^2+\alpha_s.
\label{freq2}
\end{equation}

As we can observe, the eigenvalue equation (\ref{eigen4}) involves a singular delta Dirac potential. We emphasize that it would be very interesting to analyze equation (\ref{eigen4}) using the methods outlined in references \cite{livromodelossoluveis, PRA66, JMAMG26, livroperturbacaosingular}, which are based on the boundary conditions of the associated wave functions \cite{andamento}. In this appendix, we refrain from pursuing this approach and instead leave it as an open question to be addressed in future work.

Given that the negative eigenvalues $\alpha_s$ in (\ref{eigen4}) render the Hamiltonian unbounded from below, our focus is on this particular case.

To search for possible negative eigenvalues $\alpha_s <0$, we consider solutions for $z<a$ and for $z>a$. From (\ref{eigen4}) we get respectively
\begin{eqnarray}
v_{<}(z)&=&Ae^{\kappa (z-a)},~~z<a\nonumber\\
v_{>}(z)&=&Ae^{-\kappa(z-a)},~~z>a
\label{eigenfunc}
\end{eqnarray}
where $\kappa=\sqrt{-\alpha_s}$. Note the same factor $A$ in above expressions to meet continuity of the solutions in $z=a$. Next, integrating (\ref{eigen4}), from $z=a-\epsilon$ to $z=a+\epsilon$ and taking $\epsilon\to 0$, we get
\begin{equation}
v_{<}'(a)-v_{>}'(a)+i\frac{\lambda}{2}\left[v'_{>}(a)+v'_{<}(a)\right]-i\frac{\lambda}{4}\left[v'_{>}(a)+v'_{<}(a)\right]=0
\label{quanti}
\end{equation}
where we used
\begin{equation}
\int_{a-\epsilon}^{a+\epsilon}\delta(z-a) v'(z) dz=\frac{1}{2}\left[v'_{>}(a)+v'_{<}(a)\right]
\end{equation}
since $v'(z)={dv}/{dz}$ is discontinuous at $z=a$. Using (\ref{eigenfunc}) in (\ref{quanti}) we have 
\begin{equation}
\kappa=0
\end{equation}
what contradicts the existence of negative eigenvalues $\alpha_s=-\kappa^2$. As consequence, we cannot have solutions for $\alpha_s<0$ and the hamiltonian is always bounded from below.

%%%%%%%%%

%%%%%%%%%

\end{document}